\documentclass[fleqn,10pt,twocolumn]{SICE20}
\usepackage[noend]{algpseudocode}
\usepackage{algorithm} 

\title{Modeling the Material-Inventory Transportation Problem Using Multi-Objective Optimization}

\author{Issarapong Khuankrue${}^{1\dagger}$, Sudchai Boonto${}^{1}$  and Yasuhiro Tsujimura${}^{2}$}
\speaker{Issarapong Khuankrue}

\affils{${}^{1}$Department of Control System and Instrumentation Engineering,\\ King Mongkut's University of Technology Thonburi (KMUTT), Bangkok, Thailand\\
(E-mail:l: issarapong.khu@kmutt.ac.th${}^{\dagger}$, sudchai.boo@kmutt.ac.th)\\
${}^{2}$Department of Information Technology and Media Design, \\Nippon Institute of Technology (NIT), Saitama, Japan\\
(E-mail: tujimr@nit.ac.jp)\\
}
\abstract{%
In the era of industry 4.0, the procurement in supply chain management is the key to develop information management systems. It directly affects production planning failure. In this case, it is the process to prepare and confirm the material-inventory be ordinal stages and be able to produce the products in any production line. In terms of industrial informatics, it can provide information management approaches for leveraging data sharing between factories. The multi-objective optimization will be enabled by the integration of material-inventory, production planning and monitoring, and transportation planning collaboration. The material-inventory transportation problem is the virtual factory situation when production plan failure occurs. It becomes the cost to transport material between each factory and the distribution to clients. In this study, the question of the material-inventory transportation problem is : How can we transport other materials from one factory into another factory? This study proposed a model to found out about the adjustment of material-inventory by means of transportation. The objective of this model is to minimize the whole production cost and total transportation cost.  }


\begin{document}

\maketitle


\section{Supply Chain Management in Industry 4.0}

With the development of Industry 4.0, the supply chain management is the core area of information management systems application. The procurement of inventory can directly affect production planning failure. In this case, it is the process to prepare and confirm the material-inventory be ordinal stages and be able to produce the products in any production line. 

The material-inventory transportation problem is the virtual factory situation when production plan failure occurs. It becomes the cost to transport material between each factory and the distribution to clients. In this study, the problems of inventory and transportation in factories did not have the material-inventory enough for production. The research question is : How can we transport other materials from one factory into another factory? This study proposed a way to found out about the adjustment of material-inventory by means of transportation.

\subsection{Four functions of Supply Chain Management (SCM) }
Supply Chain Management (SCM) is the way to manage the flow of products and services, including the movement and storage of raw material, inventory, and the finished goods from many points to consumption. The network of SCM is the connection and relation of channels and nodes of businesses combine in the provision of products and services required by end customers in a supply chain. The objective of SCM is to make the net value on supply with demand and measuring the performance of the production. In the industry 4.0, B. Tjahjono and others \cite{ref1} conducted their research to explore the impact of information technology or industry 4.0 on supply chain management (SCM) by focusing on four functions including 1) Procurement, 2) Transport Logistics, 3) Warehouse, and 4) Order Fulfillment.  They reviewed several sources of information and analyzed the advantages or disadvantages which impact the SCM throughout Key Performance Indicators (KPI) of each scope. The KPIs relied on the majority activities in each function. 

The procurement was categorized in purchasing activity, so the KPIs would be related to the quality of suppliers’ services and goods. In terms of the warehouse, the storage was set as KPIs to evaluate the quality of stock management. The inventories moving activity from one place to another place was considered as transport logistics’ KPIs. Finally, the fulfillment process was measured by sale activity because the quality of fulfillment in distribution channels would directly impact to sales . They found that the order fulfillment would be impacted by positive from industry 4.0 trend. Similarly, technology would be an opportunity for transport logistics. On the other side, warehouse and procurement functions were considered as opportunities or threats. They mentioned that some technology could be both opportunities and threats because all supply chain functions were related. Therefore, it depended on how to implement technology and management. Moreover, They believed that the core advantages of technology application were about increased flexibility, quality standards, efficiency, and productivity. 

\subsection{Servitization : Outcome as a Service}
Servitization concept is about customer-centric to add value through products, service, or other processes. \cite{ref2} It brings to the question:  How the production industry adapt to the big data era and suggested some essential technology for long-lasting innovative service? J. Lee and others  \cite{ref3}  reviewed that one of the significant trend which drives the manufacturing industry to be innovative as servitization.  Another factor that drove industry was a big data environment. Cyber-Physical System (CPS) framework was applied in the big data era for supporting the machines to do self-evaluation and make intelligent decisions. Moreover, they concluded that there were four areas that generated the 4.0 industry revolution.  They were prognostics and health management, SCM and business management, minimize cost, and improve the working environment. 

In the same way, J. C. De Man and J. O. Strandhagen \cite{ref4} studied the effect of 4.0 industry evolution to a sustainable business model. The descriptive meaning of a sustainable business model that it was the model transform inputs to outputs that were efficient, sufficient, and consistency delivery these values products or services to customers. However, the sustainable strategy was not applied for the long-lasting product but it used for sustainable economic advantage. They also mentioned that even the challenge of industry 4.0 in a sustainable business model that how well value proposition of business was delivered to the customer, the information technology was still able to create a competitive advantage for the business. The paper also identified that the company trended to concern about collecting data for the supply chain management.

\section{Material-Inventory Transportation Model}

\subsection{Problem Statement}
\begin{figure}[h]
	\begin{center}
		\includegraphics[width=8cm]{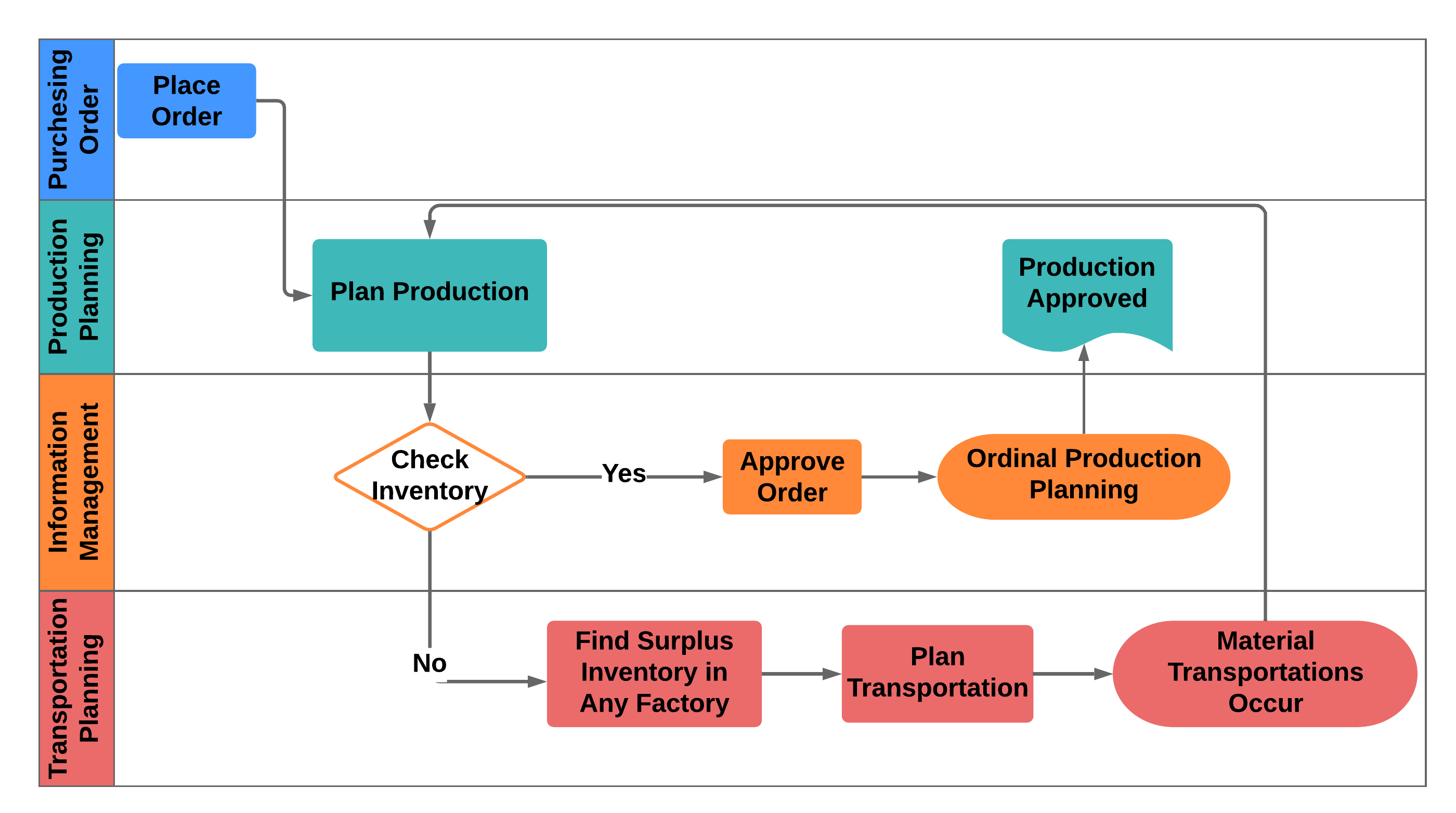}
		\caption{\label{overview} Overview of Material-Inventory Transportation Model.}
	\end{center}
\end{figure}

The Material-Inventory Transportation (MIT) problem is the virtual factory situation when production plan failure occurs. It becomes the cost to transport material ($TC$) between each factory and the distribution to clients.  As shown in Fig.\ref{overview}, some of the factories ($SI_i$)  have manufactured the same products ($y_i$) with different productivity. 

The process begins when the total order ($Y$) comes to places a purchasing order. Then in the production planning phase, an order will consider making the production plan. The information management system will support to check the necessary material ($l_k$) to make an ordered product. The information management system has to check another’s factories’ whether or not the factories have surplus inventories.  In case that the factories cannot produce the assigned amount of products due to a lack of materials, the system will make the transportation plan at the transportation planning phase.  Material transportation will occur in this state. Otherwise, If a factory has a surplus of the materials, the factory would provide the surplus to other factories, which is short of the material, with enough amount of material by transportation using a track. Ordinal production planning will be approved in this state. 

Note that, the lacks and surpluses of materials are observed by sensors embedded in the factories. The sensors, or CPS, send inventory information to the central control center through a cloud computing environment. The central control center draws up production plans for every factory and transportation plan in the virtual factory. 

In terms of SCM, inventory transportation has many logistic problems in their business and still cannot find the best way to solve the logistic problem. Transportation topics still have problems, which need to figure out and improve. This study desired to be a part of the decreasing cost of transportation. When inventory transportation occurs whenever the condition is satisfied. When the total amount of materials are over the capacity of a track, the materials are transported by enough number of tracks.
\subsection{Multi-Objective MIT Model }
In general, t10he multi-objective optimization\cite{ref5} is related to evolutionary algorithms, which is solving problems that rely on their population. It generates several elements of the optimal set, or the Vilfredo Pareto's optimal set, in a single run. Normally, the population is randomly generated to be a potential solution set. The individual, which is each solution in population, will encode and contains all the decision variables of the problem to be solved. So, the fitness function will define as the performance measurement for each of the solutions, and find the best solution when compared to the others. On the fitness function, the objective function aims to optimize. It means a variation of the objective function needs to solve the problem. Then, the stochastic selection process, or mating of individual, is the process for contributing each individual to be a higher probability of being selected. The offspring, or mutate, are generated and be the new population to be evaluated at the next iteration, or generation. 

The evolutionary algorithms are very effective in solving single-objective optimization problems. The ideal of the multi-objective optimization procedure simplified into two steps. The first step is to find multiple trade-off optimal solutions with a wide range of values for objectives. Then, choose one of the obtained solutions using higher-level information. \cite{ref6} 

The multi-objective optimization procedure will be enabled by the integration of material-inventory, production planning and monitoring, and collaboration of transportation planning. 

In this study, the main notation used in the following equations are including, $TC$ is the total transportation cost,  $PC$ is the total production cost, $SI$ is a set of factories, $l_k$ is a necessary amount of material $k$ to make a product (k =1,2, 3, …., m), $m$ is the number of kinds of materials, $L_i{}_k$ is an inventory of material $k$ at a factory $i$, $y_k$ is a set of divided amount to produce the order $Y$ by use material $k$, $q_j{}_i{}_k$ is the amount of material $k$ transported from factory $j$ to factory $i$, $r$ is the maximum number available trucks, $tc$ is a unit transportation cost, $V$ is the maximum loading capacity of a truck, and $SL$ is a set of material-inventory that need to transport which is $\Delta_i{}_j $ = 1 if some materials are transported from factory $i$ to factory $j$ or otherwise $\Delta_i{}_j$ = 0.

In the initialization stage, the defined assumption is as follows: The order and inventory in each factory are constant. Then, All trucks have the same specifications, and lastly, a transportation cost does not change according to the weight of materials. The objective of the proposed model is to minimize the production cost and transpiration cost, which can be expresses as:

\begin{eqnarray}
\begin{array}{rcl}
\textbf{Objective} :  Minimize (PC + TC) 
\end{array}
\label{eq:ss}
\end{eqnarray}

In term of constraining, The feasibility condition for the total material-inventory is
\begin{eqnarray}
\begin{array}{rcl}
\sum_{i\in I} L_{ik} \geq Y*l_{k} \mid \forall k \\
\end{array}
\end{eqnarray}

To check the occurrence condition of material-inventory transportation, The Ordinal Production Planning stage will be happened If $ (l_{ik}*y_{k} \leq L_{ik} \mid \forall i,k ) $ Then $ \Delta  = 0 $. Otherwise, the Material Transportation occur stage will be replace the ordinal stage Else If  $(l_{ik}*y_{k} > L_{ik} \mid \exists i,k) $ Then $ \Delta  = 1 $

At this step, the material-inventory transportation adjustment can be expressed in Algorithm1: CheckInventory.

\begin{algorithm}
	\caption{CheckInventory}
	\begin{algorithmic}[1]
		\State {Input: $l_k$ : a necessary amount of material k to make a product (k =1,2, 3, …., m)\\
			$m$ : the number of kinds of materials \\
			$L_i{}_k$ : an inventory of material k at a factory i \\
			$y_k$ : a set of divided amount to produce the order Y by use material k 
		}
		\State {Output: a set of of material k in $\cup^{m}_{k=1}SI_k$ }
		\State
		\While {The amount of the material k in the $factory_i$  is not enough to make $y_i$ products. }
		\State {$j \leftarrow 0$}
		\If {$l_k * y_k<=L_i{}_k$}
		\State {$\Delta \leftarrow 0$} (Ordinal Production Planning)
		\ElsIf{}
		\State $\Delta \leftarrow 1$ (Material Transportations occur)
		\EndIf 
		\For{$k \leftarrow 1$ to $ m$}
		\For{$i \leftarrow 1$ to $len(SI)$}
		\If{$l_k * y_k > L_i{}_k$ and $l_k * y_k < L_j{}_k$}
		\State $SI_k \leftarrow l_k$ 
		\EndIf
		\EndFor
		\EndFor
		\EndWhile 
		\State {return \texttt{ $\cup^{m}_{k=1}SI_k$ }}
	\end{algorithmic}
\end{algorithm}

To make the transportation plan, a constraint for the maximum number of trucks is
\begin{eqnarray}
\begin{array}{rcl}
\sum_{i\in SI_{ALL}} \{ \frac{\lceil \sum_{k\in SM_{i}} q_{ijk} \mid j \in SL_{k}, i\neq j \rceil}{V} \} \leq r
\end{array}
\end{eqnarray}

Then, The total transportation cost is \\
$TC= \sum_{i\in I} \{ \sum_{j\in SL_{k}}  \sum_{i\in SI_{ALL}} \{ \frac{\lceil \sum_{k\in SM_{i}} q_i{}_j{}_k \mid j \in SL_{k}, i\neq j \rceil}{V} \} *tc \}$

Lastly, The total production cost is $PC = \sum_{i\in I} pc_{i} * y_{i}$ .

The above model will be considered to make the decision on optimal solutions based on the objective, as the first step for multi-objective optimization. It can be do the iteration and change the wide range of values for objectives rely on the obtained solutions using higher-level information. The other factors have to concern include the type of order, production and transportation cost, and so on. 

\section{Conclusion}

To answer the question: How can we transport other materials from one factory into another factory? This paper proposed Material-Inventory Transportation (MIT) model to check and judge the number of material from one point to another point in their production environment.  It is a part of the procurement function in supply chain management. The objective of this model has been set for minimizing all production costs and total transportation costs. In this model, the information management system provides the process to prepare and confirm the material-inventory be ordinal stages and be able to produce the products in any production line, which is the data sharing between factories. This model design based on multi-objective optimization, which integrates the material-inventory, production planning, and monitoring, and collaboration of transportation planning. 

This preliminary model still on the process to develop agent-based modeling and simulation. It is able to contribute to the landscape of data, which is the possible data on the situation. It requires setting the initial behavior of the agent, depending on the business goal, and focusing on predicting the possible failure. The most feasible approach is to understand in deep of system complexity. So the new area or other functions on supply chain management could be considered on the opportunity of an information management system. The future model could also move to collect the data based on the business and supply chain experience, especially industry 4.0.


\end{document}